\documentclass[12pt]{article} 

\usepackage{latexsym}

\font\tenmsbm=msbm10 scaled 1200
\font\sevenmsbm=msbm9
\newfam\msbmfam
\textfont\msbmfam=\tenmsbm
\scriptfont\msbmfam=\sevenmsbm

\makeatletter
\@addtoreset{equation}{section}
\makeatother

\newcommand{\eref}[1]{(\ref{#1})}

\def\beq{\begin{equation}}
\def\eeq{\end{equation}}
\def\bea{\begin{eqnarray}}
\def\eea{\end{eqnarray}}
\def\bet{\begin{tabular}}
\def\eet{\end{tabular}}
\def\pa{\partial}

\def\ve{\varepsilon}
\def\qua{\quadratello}

\catcode`@=11
\def\quad@rato#1#2{{\vcenter{\vbox{
        \hrule height#2pt
        \hbox{\vrule width#2pt height#1pt \kern#1pt \vrule width#2pt}
        \hrule height#2pt} }}}
\def\quadratello{\mathchoice
\quad@rato5{.5}\quad@rato5{.5}\quad@rato{3.5}{.35}\quad@rato{2.5}{.25} }

\begin{document}
\begin{titlepage}

\begin{flushright}
Preprint DFPD 00/TH/14\\
March 2000\\
\end{flushright}

\vspace{1truecm}

\begin{center}

{\Large \bf A Quantum field theory of dyons}
\footnote{Talk given at the conference ``Quantum aspects of gauge 
theories, supersymmetry and unification", Paris, September 1999,
in collaboration with Pieralberto Marchetti.}

\vspace{2cm}

K. Lechner \footnote{kurt.lechner@pd.infn.it}
\vspace{2cm}        

{\it Dipartimento di Fisica, Universit\`a di Padova,

\smallskip

and

\smallskip

Istituto         Nazionale di Fisica Nucleare, Sezione di Padova, 
        Via F. Marzolo 8, 35131 Padova, Italy
}

\vspace{1cm}

\begin{abstract}
\vspace{0.5cm}

          We construct a classical field theory action which upon 
          quantization via the functional integral approach, gives 
          rise to a consistent Dirac--string independent quantum field
          theory. The approach entails a systematic derivation of the 
          correlators of all gauge invariant observables, and also
          of charged dyonic fields. Manifest $SO(2)$--duality invariance and
          Lorentz invariance are ensured by the PST--approach.
\end{abstract}

\end{center}
\vskip 0.5truecm 
\noindent

\end{titlepage}
\newpage
\baselineskip 6 mm

\section{Introduction}

Each formulation of a quantum field theory of dyons, particles which 
carry electric and magnetic charge, in four dimensions has to cope
with a fundamental problem: there is no natural or consistent classical
{\it field theory} action to start with. Nevertheless, there exists a 
consistent quantum field theory \cite{ZW2}.

The pathologies of the classical field theory action can be traded in 
several ways. One can renounce to describe the dynamics of the charged 
matter in terms of scalar or spinor fields \cite{IENGO1}, or one sacrifices
Lorentz--invariance introducing in the kinetic term for the gauge fields
a constant four--vector $n^\mu$ \cite{ZW2}. In the first case the quantum
field theory can not be based on a functional integral and the 
identification of the correct field strength is an open problem. In the 
second case the presence of the vector $n^\mu$ obscures the 
Lorentz--symmetry structure of the theory at the classical level
(it breaks it explicitly), and one has to show that the {\it quantum}
theory is independent of $n^\mu$ if the Dirac--Schwinger quantization
condition \cite{quant}
\beq
{1\over 2}\left(e_rg_s-e_sg_r\right)=2\pi n_{rs} 
\label{dir}
\eeq
holds, with $n_{rs}$ integer. Here $e_r(g_r)$ is the electric (magnetic)
charge of the $r$--th particle \footnote{We deal here only with the 
$SO(2)$--duality invariant theory for which \eref{dir} is the appropriate
quantization condition. For a discussion of the theory which is only
invariant under the discrete duality group $Z_4$, see \cite{PAML}. In that
case the appropriate quantization condition is Dirac's original one
$e_rg_s=2\pi n_{rs}$.}. Hence, only a posteriori $n^\mu$ acquires the 
meaning of the direction of the Dirac--string.
 
The approach we present here relies on a classical field theory
action which is manifestly invariant under Lorentz transformations
and $SO(2)$ duality, but depends on a fixed external classical 
vector field $U^\mu(x)$. The meaning of this vector field is very
simple: the unique integral curve associated to $U^\mu$ starting
from a point of a particle's trajectory determines the Dirac--string attached
to the particle in that instant. The set of all these integral curves
determines then a two--dimensional surface whose boundary is the 
trajectory of the particle. This idea can be extended to the case when
the currents are not {\it point}--{\it like}, as in the classical 
point--particle theory, but {\it continuously} distributed, as in the 
classical field theory.

The consistency check of the construction consists then in showing that
the quantum field theory, obtained from the classical field
theory action via the traditional functional integral approach,
gives rise to correlators which are independent of $U^\mu$, if 
\eref{dir} holds. Below we give the outline of the construction, 
and the proof that the partition function is indeed $U$--independent.
For correlators of generic observables, and for further developments and
details we refer the reader to \cite{PAML}.

\section{A set of equations of motion}

We start by searching for an appropriate set of equations of motion
describing the interaction of a gauge field interacting with a certain
number $N$ of dyons, with  masses $m_r$ and charges $e_r^I\equiv
(e_r,-g_r)$; $I=1,2$, $r=1,\ldots,N$. For simplicity we consider bosonic
dyons, described by complex scalar fields $\varphi_r$. To implement 
$SO(2)$--invariance in a manifest way we introduce for the photon
a doublet of one--forms $A^I$, such that the covariant derivative on
the scalars is given by 
\beq
\label{cov}
D_\mu(A) \varphi_r=\left(\pa_\mu+ie^I_r\,\ve^{IJ}A^J_\mu\right)
\varphi_r.
\eeq
The hodge duals of the total electric and magnetic currents, a doublet 
of three--forms, can then be written as
\beq
J^I={1\over 3!}dx^\rho dx^\nu dx^\mu \ve_{\mu\nu\rho\alpha}
\sum_r\,e^I_r\, i\bar\varphi_rD^\alpha\varphi_r +c.c.
\eeq
In the language of differential forms current conservation reads then simply 
\beq
\label{cons}
dJ^I=0.
\eeq
These equations allow in turn to introduce a doublet of two--forms
$C^I$ satisfying
\beq
\label{str}
J^I=dC^I.
\eeq
Clearly these forms are determined only modulo exact forms (we suppose here
to work in a four--dimensional space--time with trivial topology).

Maxwell's equations, in the presence of magnetic currents, read in this 
language 
\bea
\nonumber
dF^I &=& J^I\\
F^I&=&*\ve^{IJ}F^J,
\label{max}
\eea
where $*$ indicates the hodge dual and $\ve^{IJ}$ is the two--dimensional
antisymmetric $SO(2)$--invariant tensor. $F^2$ is the standard field 
strength two--form, and on-shell we have $F^1=*F^2$. These equations,
together with \eref{str}, allow finally to relate $F^I$ and
$C^I$ to the vector potentials
$$
F^I=dA^I+C^I.
$$

One could now close the dynamics of the system by adding just
the covariant Klein--Gordon equation for the matter fields. This is,
however, not sufficient because the fields $C^I$ are  
determined only modulo exact forms, see \eref{str}. 

A convenient
way to close the system is represented by the introduction of a vector
field $U=U^\mu(x)\pa_\mu$. We will use the
same symbol to indicate the associated one--form $U=dx^\mu U_\mu$, since
no confusion should arise.
If we indicate with $i_U$ the interior product of a form with the
vector field $U$ the supplementary condition on $C^I$ can be
written as 
\beq 
\label{supp}
i_UC^I=0,
\eeq
with the boundary condition that the fields $C^I$ vanish as $x$ goes
to minus infinity along the integral curves of $U$. It can then be
shown that the system \eref{str}, \eref{supp} admits a unique solution.
For example, if we take a constant vector $U^\mu=N^\mu$, the unique
solution can be written as \footnote{Use the identity $i_Nd+di_N=\pa_N$.}
$$
C^I={1\over \pa_N}\,i_N J^I,
$$
where $\pa_N=N^\mu\pa_\mu$. The inverse operator ${1\over \pa_N}$
has to cope with the above boundary condition and is defined by the Kernel 
$G(x)=\Theta(x_N)\delta^3\,(\vec x_N^\bot)$, $\pa_N G(x)=\delta^4(x)$,
where $\vec x_N^\bot$ are the three coordinates orthogonal to 
$x_N=x^\mu n_\mu$, and $\Theta$ is the step--function. 

We collect here the closed system of equations of motion for the fields
$A^I,C^I,\varphi_r$:
\bea
\label{1}
\left(D^\mu D_\mu +m_r^2\right)\varphi_r&=&0\\
\label{2}
F^I&=&*\,\ve^{IJ} F^J\\
\label{3}
dC^I&=&  J^I\\
\label{4}i_UC^I&=&0.
\eea
This system is manifestly $SO(2)$-- and Lorentz--invariant, but depends
on an external vector field and is therefore inconsistent. Nevertheless,
we can write an invariant action which gives rise to this system. 

Before doing that let us briefly comment on the point--particle version 
of the above system. In that case the Klein--Gordon equation is substituted by 
the generalized Lorentz--force law
$$
m_r{du_r^\mu\over d\tau_r}=\left(e_{rI}\,\ve^{IJ}\,F_J^{\mu\nu}\right)
                         (x_r)\, u_{r\nu},
$$
and the three--forms $J^I$ become a sum of $\delta$--functions along
the particle's trajectories; more precisely, we have $J^I\rightarrow
{\cal J}^I=\sum_re_r^I
{\cal J}_r$, where the three--forms ${\cal J}_r$ are the Poincar\`e--duals 
in the space
of distributional forms ($p$--``currents") of the closed 
particle's trajectories $\gamma_r$. Such $p$--forms, which are 
$\delta$--functions on a $(D-p)$--dimensional submanifold, 
are called {\it integer} forms \cite{deRham}. An important property of such 
forms
is that the integral over all space of the product of two of them 
is always an integer -- hence the name -- counting the intersection 
points of the two manifolds.

The remaining equations remain the same. In particular the solution 
for the forms $C^I$ becomes now $C^I=\sum_r e_r^I C_r$, where the
two--forms $C_r$, with $dC_r={\cal J}_r$,
are $\delta$--functions on two--dimensional surfaces
whose boundaries are the trajectories $\gamma_r$. The $r$--th surface
is composed of the integral curves of $U$ which start from the points of the 
trajectory $\gamma_r$. The boundary condition introduced above 
lets the integral curves just end on the trajectories. Therefore, in this
case the forms $C_r$ represent precisely the Dirac--string "evolving" in 
time. The Dirac--strings do not really "evolve" since they are completely
fixed by the currents, once one has chosen a vector field $U$. 

If the currents are continuously distributed, as in the field theory, also
the forms $C_r$ are spread out, but they are again uniquely determined
by the above equations.

\section{Classical field theory action}

We write now an action which gives rise to the system \eref{1}--\eref{4}. 
The scalars need the ordinary covariant Klein--Gordon action. For the 
pseudo self--duality equation of motion for Maxwell's fields we employ
the PST--approach \cite{PST}. One introduces a scalar auxiliary field
$a$ and the one--form
$$
v={da\over \sqrt{-\pa_\rho a\, \pa ^\rho a}}\equiv dx^\mu v_\mu.
$$
The PST--action can then be written as the integral of a four--form,
\beq
\label{pst}
S_0[A,C,a]={1\over 2}\int F^I\,{\cal P}(v)^{IJ}\,F^J + dA^I \,\ve^{IJ}\,C^J.
\eeq
${\cal P}(v)$ is a symmetric operator which acts in the space of 
two--forms and on the $SO(2)$--indices as
$$
{\cal P}^{IJ}(v)=vi_v*\delta^{IJ} +\left(vi_v-{1\over 2}\right)\ve^{IJ}.
$$
We remember that the PST--symmetries ensure that $a$ is non propagating
and that the (gauge--fixed) equations of motion for $A^I$ are indeed
\eref{2}. 

The equations \eref{3} and \eref{4} are implied by a 
convenient set of (auxiliary) Lagrange multiplier fields.
We introduce a doublet of auxiliary one--forms
$\tilde A^I$ and a doublet of auxiliary two--forms $\tilde C^I$. The 
action, which depends also on the fixed vector $U$,
can then be written as ($\phi\equiv(A,C,\tilde A,\tilde C,\varphi_r,a$)
\bea
\label{action}
S_U[\phi]&=&S_0
-\sum_r\int d^4x\,\bar\varphi_r(D^2_r(\tilde A)
+m_r^2)\varphi_r
\nonumber\\
&&+\int\left(\tilde A^I\ve^{IJ}dC^J-{1\over 2}\tilde C^IUi_U
\ve^{IJ}C^J
\right).
\nonumber\\
&&
\eea
Notice that in the covariant derivative for the scalars we replaced
$A^I$ with $\tilde A^I$. So, variation with respect to $\tilde A^I$
gives \eref{3}, while variation with respect to $\tilde C^I$ gives 
\eref{4}. As shown in detail in \cite{PAML}, the symmetry structure
of \eref{action} together with the equations of motion for $C^I$ 
determine also the Lagrange multiplier fields,
$$
\tilde A^I=A^I,\,\,\tilde C^I=C^I.  
$$
So there are no unwanted propagating degrees of freedom, and the 
action \eref{action} reproduces \eref{1}--\eref{4}.

\section{A representation for the partition function}

The quantum field theory can be based in a traditional manner
on the action \eref{action} through the functional integral approach.
The correlation functions of (gauge)--invariant operators are
expressed as
$$
\left\langle \,T\,O_1\cdots O_n\right\rangle
=\int \{{\cal D}\phi\}\, e^{iS_U[\phi]}\,O_1\cdots O_n,
$$
where in the functional integral measure gauge fixings of the relevant
invariances, PST--symmetries and $U(1)$--symmetries are understood, and
the integration over the fields $C^I$ inherits the boundary condition
along $U$ from the classical field theory.
As they stand, these correlation functions depend on $U$. The 
fundamental point is that one can show
that the correlation functions of all invariant operators are independent
of $U$ \cite{PAML}, if the Dirac--Schwinger quantization condition 
\eref{dir} holds. Here we limit ourselves to show that the partition
function is invariant, in that the strategy followed in the proof extends 
rather directly to the case of generic observables. To do this, we need
a convenient representation for the partition function. The rest of this
section is devoted to derive this representation, see \eref{part}.

We begin by performing the functional integration over the fields 
$A$ and $a$ which appear only in the PST--action. Since $a$
is auxiliary it can be gauge--fixed to an arbitrary function $a_0(x)$
by inserting a $\delta$--function $\delta(a-a_0)$. Clearly, the partition 
function has to be independent of $a_0$. The integration
over the fields $A^I$ is gaussian, but one has to carefully fix
the PST-- and $U(1)$--symmetries. The resulting effective action 
$\Gamma[C]$, which depends only on $C^I$, can be computed to be
\bea
\label{gamma}
e^{i\Gamma[C]}&\equiv&\int \{ {\cal D} A\}\,\{ {\cal D}a\}\,
               \,e^{iS_0[A,C,a]}\\  
\Gamma[C]&=& -{1\over 2}\int\left(dC^I{*\over\quadratello}dC^I
             -dC^I{\ve^{IJ}\over\quadratello}*d* C^J\right)
\nonumber
\eea
Notice that it depends on the currents not only through $J^I=dC^I$, but
also through the ``Dirac--strings" $C^I$ in the second term.

The integration over the scalar fields amounts to the evaluation
of the corresponding covariant Klein--Gordon determinants. These 
determinants can be represented in a standard way as Feynman 
{\it path}--integrals over {\it classical particles trajectories} 
\cite{path}. In the case at hand, for the product of all the 
determinants the relevant representation can be written in a compact
way as
\beq
\prod_{r=1}^N {\rm det}^{-1}
\left(-i\left(D^2_r(\tilde A)+m_r^2\right)\right)=
\int\{{\cal D}\gamma\}e^{-i\sum_{r=1}^N \oint_{\gamma_r} 
\tilde A^I\,\ve^{IJ}\,e_r^J}.
\label{feyn}
\eeq
Here $\{{\cal D}\gamma\}$ indicates a (complicated) measure over
the closed particle's trajectories $\gamma_r$ whose details are, 
however, not needed for our purposes. The essential feature of
this representation is that it retrieves in the quantum field theory
a classical point particle nature. This point is essential for what
concerns the proof of Dirac--string independence, as we will see in
a moment. The exponent in this representation can also be rewritten as
the integral of a four--form 
$$
\sum_{r=1}^N \oint_{\gamma_r}\tilde A^I\,\ve^{IJ}\,e_r^J=
\int \tilde A^I\ve^{IJ}{\cal J}^J,
$$
where ${\cal J}^I=\sum_r e^I_r {\cal J}_r$ and the three--forms 
${\cal J}_r$ are just $\delta$--functions on the curves $\gamma_r$, i.e.
their Poincar\`e--duals. Since these curves are closed we have 
$d{\cal J}^I=0=d{\cal J}_r$.

Integration over $\tilde C^I$ gives the 
$\delta$--functions
\par\noindent $\delta(i_UC)$ and, suppressing the $SO(2)$--indices,
the partition function can be written as
\bea
Z&=&\int \{{\cal D}\phi\}\, e^{iS_U[\phi]}\nonumber\\
&=&\int\{{\cal D}\gamma{\cal D}C{\cal D}\tilde A\}\delta(i_UC)
e^{i\Gamma[C]+i\int\tilde A\ve(dC-{\cal J})}
\nonumber\\ 
\nonumber
&=& \int\{{\cal D}\gamma{\cal D}C\} \delta \left(i_UC\right)
\delta\left(dC-{\cal J}\right)\,e^{i\Gamma[C]}. 
\eea
The $\delta$--functions restrict the variables 
$C^I$ to
\bea\label{curr}
dC^I&=&{\cal J}^I\\
\label{const}
i_UC^I&=&0,
\eea
which, are precisely the classical
field theory equations for these fields, but now
the currents correspond to classical {\it point}--{\it like} particles. 
Taking the boundary
conditions into account, we know that these equations admit a unique 
solution,
\beq
C^I(U)=\sum_r e^I_r C_r(U),
\eeq
which depends on $U^\mu$ and on the classical currents ${\cal J}_r$.
This allows eventually to write the partition function as 
\beq
\label{part}
Z=\int\{{\cal D}\gamma\} \,e^{i\Gamma[C(U)]}. 
\eeq
Using this representation for $Z$ we can now show that it is 
$U$--independent. 

\section{Dirac--anomaly and Dirac--string independence} 

We recall that the two--forms $C_r(U)$ are $\delta$--functions on the
surfaces made out of integral curves of $U$ ending on the trajectories
$\gamma_r$. $Z$ depends on $U$ only through $C_r(U)$.
Under a (finite) change of $U^\mu\rightarrow U'^\mu$ the two--forms $C_r(U)$
change by an exact form
$$
C_r(U')=C_r(U)+dH_r(U',U),
$$
as can be seen from \eref{curr}, (the currents ${\cal J}$ remain clearly
fixed). The important point is, however, that the one--forms $H_r$ are 
{\it integer} forms, they are $\delta$--functions on a three--manifold,
which is bounded by the two two--dimensional surfaces associated to 
$C_r(U)$ and $C_r(U')$. Therefore, changing $U$ amounts precisely to change
the Dirac--string attached to the particle in each point: it moves from 
an integral curve of $U$ to an integral curve of $U'$. 
For the total $C^I$ we have then 
$$
C'^I=C^I+dH^I,\,\,\, H^I\equiv \sum_r e^I_r H_r.
$$
We can now evaluate the "Dirac--anomaly'', i.e. the variation of the effective
action $\Gamma[C(U)]$, see \eref{gamma}, under this change. Since the $C^I$
change by an exact differential, only the second term in \eref{gamma}
contributes and one gets for the 
Dirac--anomaly\footnote{Use $*d*d+d*d*=\qua$.}
\bea 
{\cal A}_D&\equiv& \Gamma[C(U')]-\Gamma[C(U)]\nonumber\\
          &=&{1\over 2}\int {\cal J}^I\ve^{IJ}H^I\nonumber\\
          &=& {1\over 2}\sum_{r,s}\,(e_r^I\,\ve^{IJ}\, e_s^J)
           \int {\cal J}_r H_s\nonumber\\
          &=& 2\pi \sum_{r,s}n_{rs}\int {\cal J}_r H_s,\label{anom}
\eea
where in the last line we used the charge quantization condition
\eref{dir}. Since the three--forms ${\cal J}_r$ as well as the one--forms
$H_s$ are integer forms, also the integrals in the last line are integer,
and the Dirac--anomaly becomes an integer multiple of $2\pi$. Therefore, 
under a change of $U$ the exponent in \eref{part} changes by an integer 
multiple of $2\pi$ and the partition function is $U$--independent.

\section{Further developments and concluding remarks}

The strategy illustrated above can be generalized to prove 
Dirac--string independence of the correlators of generic observables.
It extends in a straightforward way to the correlators of
currents, Wilson loops, and neutral Mandelstam--string observables 
(''mesons"). The correlators of the electromagnetic field strength 
$F^I=(F^1,F^2)$ and of {\it charged} operators, instead, present
additional problems.

The difficulty with the electromagnetic field strength is related with
the fact that, to cope with manifest duality, we have introduced two 
of them, $F^1$ and $F^2$. Only if the classical equations of motion
\eref{2} hold we have the identification $F^1=*F^2$, but this relation
does not hold in the functional integral. This means that the correlators
of $F^1$ do not coincide with the correlators of $*F^2$. This mismatch
is solved by the observation that the quantities $F^I$ can not represent
the electromagnetic field strength {\it off--shell}. They are, in fact, 
gauge invariant, but they are not invariant under the PST--symmetries.
We need a couple of two--forms $K^I$ which are invariant under the 
PST--symmetries and which reduce on--shell (i.e. under \eref{2}) to
the $F^I$. Such forms exist, indeed, and they are given by 
$$
K^I=F^I-vi_v\left(F^I-*\ve^{IJ}F^J\right).
$$
The key point is that the fields $K^I$ satisfy, moreover, identically the
pseudo self--duality relation
$$
K^I=*\ve^{IJ}K^J.
$$
Their correlators solve, therefore, automatically the problem related with
the mismatch between $F^1$ and $*F^2$. It can also be shown that, despite
the explicit appearance of the field $a$ in their definition, the 
correlators of the $K^I$'s are independent of $a_0(x)$, the gauge--fixed
$a$--field, manifestly Lorentz--invariant and $U$--independent.

The problem regarding the correlators of charged fields is related with 
the correct definition of the related gauge--invariant charged field
operators. The extension of Mandelstam's proposal \cite{Mandel} for such 
field operators
to the dyonic case would be given by\footnote{For an explanation of the
appearance of the fields $\tilde A^I$ instead of the fields $A^I$ see
\cite{PAML}.}
\beq
\label{man}
\phi_r(x,\gamma_x) = \varphi_r(x)\,exp
\left(i \int_{\gamma_x} e_r^I\,\ve^{IJ}\,\tilde A^J\right),
\eeq
where $\gamma_x$ indicates a path which goes from $x$ to infinity, the
Mandelstam--string. The correlators of these operators can indeed be seen
to be Dirac--string independent -- in the present formulation 
$U$--independent -- but they are plagued by (non--renormalizable)
infrared divergences, due to the infinite extension of the Mandelstam
string. 

An alternative proposal for charged operators, due to Dirac,
corresponds to substitute in \eref{man} the ``singular" 
Mandelstam--string with a radially symmetric Coulomb potential, which 
behaves as $1/r^2$, and cures the infrared divergences. But this time the
electric flux is spread out continuously in space, and the correlators
depend on the Dirac--string (the Coulomb potential is not an ``integer
form"). 

A solution of the problem has been proposed in \cite{PAM}, starting
from Mandelstam's proposal. One replaces the single Mandelstam--string
$\gamma_x$ with a sum over such strings, weighted by a convenient
measure
$$
\Phi_r(x)=
\varphi_r(x)\int{\{\cal D}\gamma_x\}\,
exp\left(i \int_{\gamma_x} e_r^I\,\ve^{IJ}\, \tilde A^J\right).
$$
The corresponding correlators are now Dirac--string independent. Moreover,
the measure ${\{\cal D}\gamma_x\}$ has been constructed (implicitly)
in \cite{PAM}, and there it has also been shown that at large distances,
on average, it reproduces the Coulomb potential.

The method presented in this talk applies equally
well to fermions; the Feynman path--integral representations for
the determinants, like
\eref{feyn}, are available for spinor fields, too. Also, the introduction
of $\vartheta$--angles does not encounter any difficulty.
Due to manifest 
Lorentz--invariance of the PST--approach, it admits also a canonical
diffeomorphism invariant 
coupling to gravity. Dirac--string independence follows in this case
from the analogous result in the flat case. This is due to the fact that
the Dirac--anomaly \eref{anom} is a topological invariant, i.e.
metric independent. 

For the duality properties of the model we
refer the reader to \cite{PAML}.

The techniques illustrated in this talk can be also applied to
a system of interacting $p$--branes, dual branes, and dyonic branes 
in a generic $D$--dimensional space--time \cite{LM}.

The formulation of a quantum field theory for dyons presented here
can be seen to be equivalent to previous formulations 
\cite{ZW2,IENGO1}, for what concerns
the details which have been worked out in those formulations. 

The advantage
of the present formulation is constituted by the manifest 
Lorentz--invariance at each step, by the clear identification of
the Dirac--string (represented by the field $U$) from the beginning, and 
by a systematic derivation of the observables which are triggered by
the symmetries of the action \eref{action}.

\paragraph{Acknowledgements.}

This work was supported by the 
European Commission TMR programme ERBFMPX-CT96-0045.

\end{document}